\documentclass [11pt]{article}
\usepackage{amssymb}
\usepackage{amsmath}
\usepackage{graphicx}

\begin{document}

\title{A Note on B\"odewadt-Hartmann Layers}
\author{P. A. Davidson, A. Poth\'erat \\
Department of Engineering, University of Cambridge}
\date{16 May, 2002}
\maketitle


\begin{abstract}
This paper addresses the problem of axisymetric rotating flows bounded by a 
fixed horizontal plate and subject to a permanent, uniform, vertical 
magnetic field (the so-called B\"odewadt-Hartmann problem). The aim is to 
find out which one of the Coriolis or the Lorentz force dominates the 
dynamics (and hence the boundary layer thickness) when their ratio, 
represented by the Elsasser number $A$, varies. After a short review of 
existing linear solutions of the semi infinite Ekman-Hartmann problem, 
weakly non-linear 
analytical solutions as well as fully non-linear numerical solutions are 
given.

The case of a rotating vortex in a finite depth fluid layer is then studied, 
first when the flow is steady under a forced rotation and second for
 spin-down from some initial state. The angular velocity in the first 
case and decay time in the second are obtained analytically as a function of 
$A$ using the weakly non linear results of the semi-infinite 
B\"{o}dewadt-Hartmann problem.


\end{abstract}

\section{Introduction}

\begin{figure}[htbp]
\centering
\includegraphics[width=0.5\textwidth]{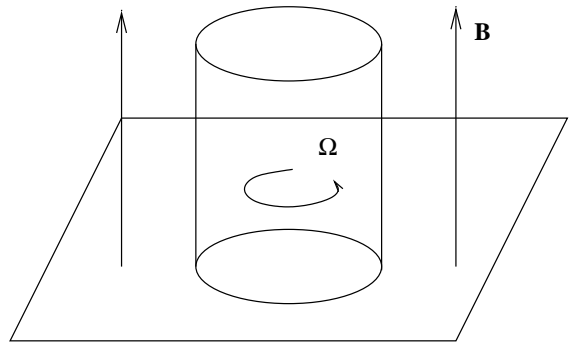}
\caption{B\"{o}dewadt-Hartmann Configuration.}
\label{bodew_ha}
\end{figure}

We are concerned here with the interaction of a vortex with a plane surface 
in the presence of an imposed magnetic field $\mathbf B$. The axis of the vortex and the 
magnetic field are both normal to the surface and, for simplicity, we take 
the flow to be axisymmetric (see Figure \ref{bodew_ha}). We are interested in 
characterizing the decay of the vortex, due either to surface friction or to 
magnetic damping. Such geometries are important in geophysics (motion in 
planetary interiors are dominated by Coriolis and Lorentz forces), in 
engineering (for example, in the magnetic damping of turbulence in 
castings), and in laboratory studies of MHD turbulence.

\begin{figure}[htbp]
\centering
\includegraphics[scale=0.6]{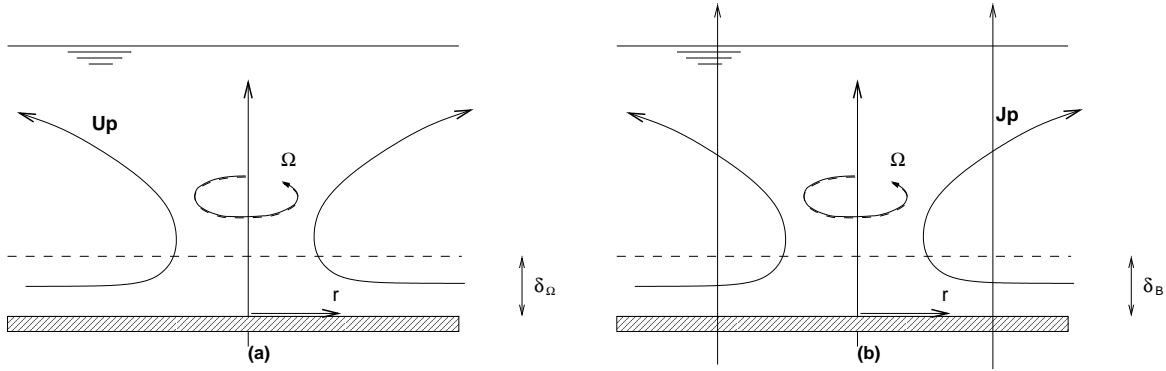}
\caption{Classical Boundary layers. (a) B\"{o}dewadt layer; (b) Hartmann layer. In 
the Hartmann layer, \textbf B is dominant in the sense that $\sigma B^2 >>\rho \Omega $.}
\label{classical_bl}
\end{figure}

Our geometry combines two classical problems in fluid mechanics: the 
B\"{o}dewadt layer and the Hartmann layer. These are illustrated in Figure 
\ref{classical_bl}. In the B\"{o}dewadt problem there is no magnetic field and the fluid is 
in a state of rigid-body rotation above a plane surface. A boundary layer 
develops, of approximate thickness,
\begin{displaymath}
\delta _\Omega = \left( {\nu / \Omega } \right)^{1 \mathord{\left/ {\vphantom 
{1 2}} \right. \kern-\nulldelimiterspace} 2} 
\end{displaymath}

\noindent
where $\nu$ is the fluid viscosity and \textit{$\Omega $} is the core rotation rate. Within this 
boundary layer there is an imbalance between the local centrifugal force, 
$\rho u_\theta ^2 / r$, and the radial pressure gradient, ${\partial p} 
\mathord{\left/ {\vphantom {{\partial p} {\partial r}}} \right. 
\kern-\nulldelimiterspace} {\partial r}$, which is established outside the 
boundary layer by the rigid-body rotation
\footnote{ We 
use cylindrical polar coordinates (\textit{r,$\theta $,z}) throughout.}.
 That is, the core 
rotation sets up a radial pressure gradient of ${\partial p} \mathord{\left/ 
{\vphantom {{\partial p} {\partial r}}} \right. \kern-\nulldelimiterspace} 
{\partial r} = \rho \Omega ^2r$ and this is imposed on the boundary layer 
where $u_{\theta }$ is locally diminished due to viscous drag. The result is 
a radial inflow within the boundary layer. By continuity there is an upward 
flux of mass out of the B\"{o}dewadt layer and into the core, and in the 
configuration shown in Figure \ref{classical_bl}(a) this leads to a weak secondary flow in 
the core, \textbf{u}$_{p}$. This secondary (poloidal) flow is crucial to the 
development of the vortex, since it sets up a Coriolis force, $ - 2u_r 
\Omega {\rm {\bf \hat {e}}}_\theta $, which opposes the core motion and 
tends to decelerate the vortex. This kind of motion is seen, for example, in 
the spin-down of a stirred cup of tea.

The Hartmann problem is shown in Figure \ref{classical_bl}(b). Here the inertial forces are 
weak and the Lorentz force is strong in the sense that the Elsasser number
\begin{equation}
\label{eq1}
A = \frac{\sigma B^2}{2\rho \Omega }
\end{equation}

\noindent
is assumed large. (\textit{$\sigma $} is the electrical conductivity of the fluid.) A boundary 
layer is again established, although this time it turns out to have a 
thickness of the order of 
\begin{equation}
\label{eq2}
\delta _B = \left( {\nu\tau } \right)^{1 \mathord{\left/ {\vphantom {1 2}} 
\right. \kern-\nulldelimiterspace} 2}\;\;\;\;,\;\;\;\;\;\tau ^{ - 1} = 
\sigma B^2 / \rho 
\end{equation}

\noindent
where \textit{$\tau $} is the so-called Joule damping time. Within this boundary layer an 
electric current flows in accordance with Ohm's law
\begin{equation}
\label{eq3}
{\rm {\bf J}} = \sigma \left( {{\rm {\bf u}}\times {\rm {\bf B}} - \nabla V} 
\right)
\end{equation}

\noindent
where  $V$ is the electrostatic 
potential. (The induced magnetic field, defined via $\nabla \times {\rm {\bf 
b}} = \mu {\rm {\bf J}}$, is neglected throughout on the assumption that 
$\mu \sigma u\delta$ is small. This is valid in laboratory and engineering applications, but is 
not always true in the core of the earth.) The electric field, ${\rm {\bf 
E}} = - \nabla V$, in the Hartmann layer is set by the electric field in the 
core flow, ${\rm {\bf E}} = - \Omega rB{\rm {\bf \hat {e}}}_r $, and this 
dominates the weaker ${\rm {\bf u}}\times {\rm {\bf B}}$ term in the 
boundary layer. The net result is radially inward flow of current. 
Continuity of current then requires that there is an upward flow of current 
out of the boundary layer and into the core. In the configuration shown in 
Figure \ref{classical_bl}(b) this leads to a weak poloidal current, $\mathbf J_{p}$, in the 
core. (Note the similarity between \textbf{u}$_{p}$ and \textbf{J}$_{p}$ in 
Figures \ref{classical_bl}(a) and \ref{classical_bl}(b).) This core current is crucial since it results in a 
Lorentz force ${\rm {\bf J}}\times {\rm {\bf B}} = - J_r B{\rm {\bf \hat 
{e}}}_\theta $ which retards the core vortex. 

\begin{figure}[htbp]
\centering
\includegraphics[width=0.75\textwidth]{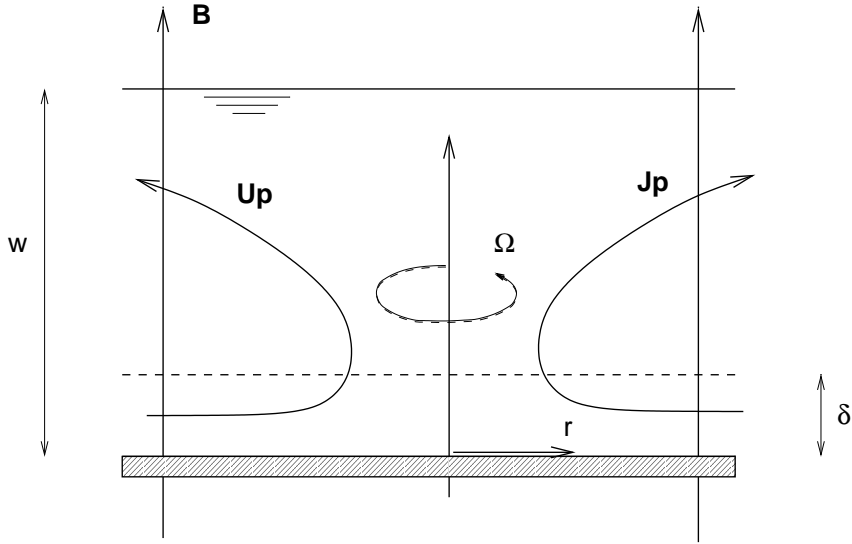}
\caption{B\"{o}dewadt-Hartman geometry. The secondary flow generates a Coriolis 
Force, $2u_r$\textit{$\Omega $, }which tends to oppose the vortex outside the boundary layer. 
The induced current interacts with \textbf{B} to create a Lorentz force 
$J_{r}B, $which also opposes the vortex.}
\label{fig3}
\end{figure}

The crucial feature of both the B\"{o}dewadt and Hartmann flows is that 
they constitute active boundary layers, in the sense that they react back on 
the core flow which created them in the first place. The problem of interest 
here is shown in Figure \ref{fig3}. The Elsasser number is allowed to be big or 
small, so that we may capture both the B\"{o}dewadt problem, when $A\to 
0$, and the Hartmann problem, $A$~$ \to $~\textit{$\infty $}. When $A\sim 1$ we expect both 
phenomena to be present. The questions which are important are: (i) how does 
the boundary layer thickness scale with $A$; and (ii) for a given value of $A$, is 
the deceleration of the core vortex due primarily to the Coriolis force, $ - 
2u_r \Omega {\rm {\bf \hat {e}}}_\theta $, or to the Lorentz force, $ - J_r 
B{\rm {\bf \hat {e}}}_\theta $?

There is close correspondence between our problem and the well-known 
Ekman-Hartmann layer. This latter is shown in Figure \ref{ekman_ha}. An Ekman layer is 
formed when there is an infinitesimal difference in rotation between a 
rapidly rotating fluid and an adjacent, plane surface. When the surface 
rotates slightly slower than the fluid we find a secondary flow very like 
that in the B\"{o}dewadt problem. Indeed, the mechanism which generates the 
poloidal flow is essentially the same as in a B\"{o}dewadt layer. Thus, 
phenomenologically, Ekman layers and B\"{o}dewadt layers are very similar. 
When a magnetic field is added to an Ekman layer we get the Ekman-Hartmann 
problem, which shares many of the same characteristics as a 
B\"{o}dewadt-Hartmann layer. However, one of the main differences is that, 
when viewed in a rotating frame of reference, inertia is negligible in the 
Ekman-Hartmann problem. (In fact, we interpret the phrase `rapid rotation' 
to mean that ${\rm {\bf u}} \cdot \nabla {\rm {\bf u}}$ is negligible by 
comparison with the Coriolis force, $2{\rm {\bf u}}\times \Omega $.) Thus 
the Ekman and Ekman-Hartmann problems are linear. The B\"{o}dewadt-Hartmann 
problem, on the other hand, is not.

The layout of the paper is as follows. In section 2 we review the properties 
of an Ekman-Hartmann layer in a semi-infinite fluid. The local properties of 
such layers are well-known. (See, for example, \cite{acheson73}.) 
However, we are interested here in axisymmetric flows of the Karman type 
($u_{r}$ and $u_{\theta }$ linear in $r)$ and so we redevelop the conventional 
analysis in cylindrical polar coordinates and restrict solutions to those 
with Karman similarity. This allows us to place the subsequent non-linear 
problem in context. Next, in section 3, we focus on B\"{o}dewadt-Hartmann 
layers. Once again, the discussion is restricted to a semi-infinite domain. 
Here we develop the ideas of \cite{stephenson69} and \cite{loffredo86}, who 
noted that such layers admit self-similar solutions of the Karman type. 
However, we go further than these authors, developing approximate solutions 
for these Karman-like flows, the validity of which is confirmed by fully 
non-linear numerical simulations.

The novel results of the paper lie in sections 4 and 5 where we move from 
semi-infinite domains to confined flows. There is then a coupling of the 
core motion to the boundary layer through the radial components of 
\textbf{u} and \textbf{J} (see Figure \ref{fig3}). The associated Lorentz and 
Coriolis forces tend to oppose the core motion and the central questions now 
relate to the influence of the boundary layer on the external vortex, rather 
than on the boundary layer itself. There are two canonical problems of 
interest here. One is the steady-state case in which the core vortex is 
maintained by some external azimuthal force (say that generated by a 
rotating magnetic field) and the other is the transient problem of spin-down 
from some initial state of rotation. In both cases we are interested in 
determining the dominant force balance in the core. (Does the primary 
resistance to motion come from the Lorentz force or the Coriolis force?) In 
the steady flow we determine the magnitude of \textit{$\Omega $} as a function of $A$, while in 
the transient problem we calculate the spin-down time, which also depends on 
$A$.

There are two dimensionless groups which appear throughout. We have already 
mentioned the Elsasser number which provides a measure of the relative sizes 
of the Lorentz and Coriolis forces,
\begin{equation}
\label{eq4}
A = \frac{\sigma B^2}{2\Omega \rho } = \frac{1}{2\Omega \tau } = 
\frac{\delta _\Omega ^2 }{2\delta _B^2 }.
\end{equation}

This usually lies in the range $0<A<10$, and very rarely exceeds 50. On the 
other hand, the Reynolds number, $Re = \Omega W^2 / \nu$ is invariably very 
large (Here $W$ is the depth of fluid: see Figure \ref{fig3}). Thus we consider the 
range of parameters: 
\begin{equation}
\label{eq5}
0 < A < < Re\;\;\;\;,\;\;\;\;\;Re > > 1.
\end{equation}

However, the flow is assumed to be laminar so that, in practice, $Re$ cannot 
be made too large.

\section{Ekman-Hartmann Layers of the Karman Type}
Ekman and Hartmann layers are usually described as a local phenomenon, the 
boundary layer being the result of some local difference in the core and 
boundary velocities. As a result, they are usually discussed in a planar 
framework, using cartesian coordinates. Since we are ultimately interested 
in axisymmetric, nonlinear flows of the Karman type, we shall take a 
different approach. We restrict ourselves to axisymmetric motion, described 
using cylindrical polar coordinates ($r$,\textit{$\theta $},$z)$, and look for Ekman-Hartmann layers 
which possess Karman similarity ($u_{r}$ and $u_{\theta }$ linear in $r$). 

\begin{figure}
\centering
\includegraphics[width=0.5\textwidth]{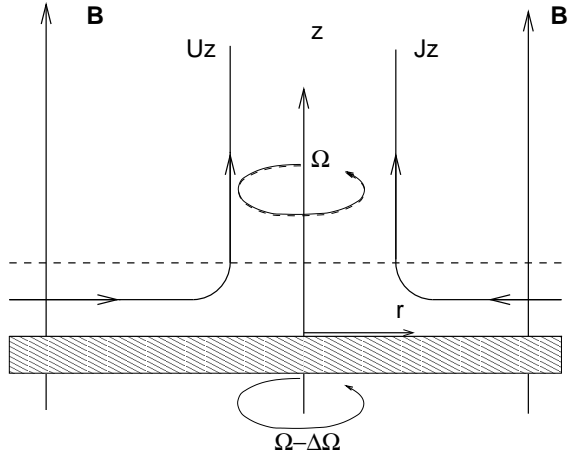}
\caption{Ekman-Hartmann geometry}
\label{ekman_ha}
\end{figure}

Consider a conducting fluid which fills the domain $z>0$ and rotates above a 
plane, insulating surface located at $z=0$ (see Figure \ref{ekman_ha}). Far from the 
surface we have rigid-body rotation, $u_\theta = \Omega r$, and the surface 
itself rotates at the lower rate,$\Omega - \Delta \Omega $, $\Delta \Omega < 
< \Omega $. Both $\Omega $ and $\Delta \Omega $ are assumed to be constant 
and so, in a frame of reference rotating with the unperturbed fluid we have,
\begin{eqnarray}
\label{eq6}
0 &=& 2{\rm {\bf u}}\times \Omega - \nabla \left( {p / \rho } \right) + 
\nu\nabla ^2{\rm {\bf u}} + {\rm {\bf J}}\times {\rm {\bf B}} / \rho \\ 
\label{eq7}
{\rm {\bf J}} &=& \sigma \left( {{\rm {\bf u}}\times {\rm {\bf B}} - \nabla V} 
\right),
\end{eqnarray}

\noindent
where \textbf{B} is a uniform, imposed magnetic field which is parallel to 
\textbf{$\Omega $}. (The inertial term, ${\rm {\bf u}} \cdot \nabla {\rm 
{\bf u}}$, is assumed to be much smaller than the Coriolis force, and so is 
omitted from (\ref{eq6}).) Taking the curl of (\ref{eq6}) twice and substituting for 
\textbf{J} yields,
\begin{equation}
\label{eq8}
\nu\nabla ^4{\rm {\bf u}} - \left( {\sigma / \rho } \right)\left( {{\rm {\bf 
B}} \cdot \nabla } \right)^2{\rm {\bf u}} = 2\left( {\Omega \cdot \nabla } 
\right)\mathbf{\omega}, 
\end{equation}

\noindent
where $\mathbf{\omega} = \nabla \times {\rm {\bf u}}$. From this we may obtain the 
governing equation for \textbf{u}:
\begin{equation}
\label{eq9}
\left[ {\nu\nabla ^4 - \frac{1}{\tau }\,\frac{\partial ^2}{\partial z^2}} 
\right]^2{\rm {\bf u}} + \left( {2\Omega \cdot \nabla } \right)^2\nabla 
^2{\rm {\bf u}} = 0
\end{equation}
\noindent
(See, for example, \cite{acheson73}). Let us now look for 
axisymmetric solutions of the Karman form:
\begin{equation}
\label{eq10}
{\rm {\bf u}} = {\rm {\bf u}}_p + {\rm {\bf u}}_\theta = - \nabla \times 
\left[ {r\Psi (z){\rm {\bf \hat {e}}}_\theta } \right] + rG(z){\rm {\bf \hat 
{e}}}_\theta .
\end{equation}
\noindent
We find that both $G$ and ${\Psi }'(z)$ satisfy,
\begin{equation}
\label{eq11}
\left[ {\frac{\delta _\Omega ^2 }{2}\,\frac{\partial ^2}{\partial z^2} - A} 
\right]^2\left( {G,{\Psi }'} \right) + \left( {G,{\Psi }'} \right) = 0,
\end{equation}

\noindent
where $\delta _\Omega$ is the B\"{o}dewadt (or Ekman) boundary-layer scale,
\begin{equation}
\label{eq12}
\delta _\Omega = \left( {\nu / \Omega } \right)^{1 \mathord{\left/ {\vphantom 
{1 2}} \right. \kern-\nulldelimiterspace} 2}.
\end{equation}

\noindent
Next we solve for $G$ and ${\Psi }'$. After a little algebra we find,
\begin{eqnarray}
\label{eq13}
\frac{u_r }{\Delta \Omega r}&=& - \exp \left[ { - Rz / \delta _\Omega } 
\right]\sin \left[ {z / R\delta _\Omega } \right]\\
\label{eq14}
\frac{u_\theta }{\Delta \Omega r} &=& - \exp \left[ { - Rz / \delta _\Omega } 
\right]\cos \left[ {z / R\delta _\Omega } \right]\\
\label{eq15}
\frac{u_z - \left( {u_z } \right)_\infty }{2\Delta \Omega \delta _\Omega } &=& 
- \exp \left[ { - Rz / \delta _\Omega } \right]\,\,\left[ {\frac{R}{1 + 
R^4}\cos \left( {\frac{z}{R\delta _\Omega }} \right) + \frac{R^3}{1 + 
R^4}\sin \left( {\frac{z}{R\delta _\Omega }} \right)} \right]\\
\label{eq16}
\left( {u_z } \right)_\infty &=& 2\Delta \Omega \delta _\Omega \frac{R}{1 + 
R^4}
\end{eqnarray}

\noindent
where $R$ is a function of the Elsasser number,
\begin{equation}
\label{eq17}
R = \left[ {A + \left( {1 + A^2} \right)^{1 \mathord{\left/ {\vphantom {1 
2}} \right. \kern-\nulldelimiterspace} 2}} \right]^{1 \mathord{\left/ 
{\vphantom {1 2}} \right. \kern-\nulldelimiterspace} 2}
\end{equation}
\noindent
Returning now to (\ref{eq7}) we evaluate \textbf{J}. In particular we find that, 
outside the boundary layer, we have,
\begin{equation}
\label{eq18}
\left( {J_z } \right)_\infty = 2\sigma B\Delta \Omega \delta _\Omega 
\frac{R^3}{1 + R^4}
\end{equation}
\noindent
It is convenient that all of the electromagnetic effects are bound up in the 
single parameter, $R$. If we let $A$~$ \to $~0 we capture the conventional Ekman 
solution,
\begin{eqnarray}
\label{eq19}
\frac{u_r }{\Delta \Omega r} = - \exp \left[ { - z / \delta _\Omega } 
\right]\sin \left[ {z / \delta _\Omega } \right] \quad &,&
\quad
\frac{u_\theta }{\Delta \Omega r} = - \exp \left[ { - z / \delta _\Omega } 
\right]\cos \left[ {z / \delta _\Omega } \right]\\
\label{eq21}
\left( {u_z } \right)_\infty &=& \Delta \Omega \delta _\Omega 
\end{eqnarray}


\noindent
Conversely, if we let $A\to\infty $ then we obtain the Hartmann 
solution,
\begin{equation}
\label{eq22}
u_r = u_z = 0,
\quad
u_\theta = - \Delta \Omega r\,\exp \left[ { - z / \delta _B } \right],
\quad
(J_z )_\infty = 2\sigma B\Delta \Omega \delta _B 
\end{equation}

Thus we have a smooth transition from a Coriolis dominated flow to a Lorentz 
dominated motion. Of particular interest is the far-field values of 
\textbf{u} and \textbf{J}, since it is these which feed into the core flow. 
In dimensionless form these are related by,
\begin{equation}
\label{eq23}
\frac{\left( {J_z } \right)_\infty }{\sigma B(u_z )_\infty }\, = A + \left( 
{1 + A^2} \right)^{1 \mathord{\left/ {\vphantom {1 2}} \right. 
\kern-\nulldelimiterspace} 2}
\end{equation}

\noindent
If we define the boundary-layer thickness, \textit{$\delta $}, to be distance over which 
$u$ declines by a factor of $e^{-1}$, then we also have,
\begin{equation}
\label{eq24}
\delta ^2 = \frac{\delta _\Omega ^2 }{\left( {1 + A^2} \right)^{1/2}+ 
A} = \frac{2\delta _B^2 }{1 + 
\left({1 + A^{ - 2}} \right)^{1/2}}
\end{equation}
\noindent
Note that when $\delta _\Omega $ and $\delta _B $ are very different there 
is still only one relevant length scale, \textit{$\delta $}. That is, there is no nesting of 
the boundary layers, with one lying within the other. Comparing (\ref{eq23}) with 
(\ref{eq24}) we see that, for arbitrary $B$,
\begin{equation}
\label{eq25}
 \frac{(J_z )_\infty }{\sigma B(u_z )_\infty } = \left( {\frac{\Omega \delta 
^2}{\nu }} \right)^{ - 1}  
\end{equation}
\noindent
The inward flow of mass and current in the boundary layer is essentially for 
the reasons given in Section 1. The mass flow arises from the radial pressure
 gradient set up in the core flow. A similar argument explains the 
radial current. Outside the boundary layer the electrostatic term in Ohm's 
law (\ref{eq7}) is balanced by $\mathbf{u_\theta} \times {\rm {\bf B}}$
\begin{displaymath}
E_r = - \frac{\partial V}{\partial r} = - u_\theta B = - \Omega Br
\end{displaymath}
\noindent
This electric field is also imposed on the boundary layer where $ - u_\theta 
B$ is insufficient to balance it. The result is a flow of current as shown 
in Figure \ref{ekman_ha}.

\section{B\"{o}dewadt-Hartmann Layers in a Semi-infinite Fluid}

\subsection{The Governing Equations}
Let us now turn to the non-linear problem in which the plate in Figure 4 is 
stationary. That is, we consider a B\"{o}dewadt-Hartmann layer in a 
semi-infinite fluid. As noted by \cite{stephenson69}, such a layer admits a 
Karman-like solution in which $u_\theta $ and $u_{r}$ are linear in $r$. This 
time our governing equations, in an absolute frame of reference, are
\begin{eqnarray}
\label{eq26}
{\rm {\bf u}} \cdot \nabla {\rm {\bf u}} = - \nabla \left( {p / \rho } 
\right) + \nu\nabla ^2{\rm {\bf u}} + {\rm {\bf J}}\times {\rm {\bf B}} / \rho\\ 
\label{eq27}
{\rm {\bf J}} = \sigma \left( {{\rm {\bf u}}\times {\rm {\bf B}} - \nabla V} 
\right)
\end{eqnarray}
\noindent
From this we find
\begin{equation}
\label{eq28}
\nu\nabla ^4{\rm {\bf u}} - \left( {\sigma / \rho } \right)\left( {{\rm {\bf 
B}} \cdot \nabla } \right)^2{\rm {\bf u}} = \nabla \times \nabla \times 
({\rm {\bf u}}\times \bf{\omega} )
\end{equation}

\noindent
which might be compared with (\ref{eq8}). We now look for Karman-like solutions of 
the form:
\begin{displaymath}
{\rm {\bf u}} = \left[ {\Omega rF(z / l),\;\;\Omega rG(z / l),\,\;\;\Omega 
lH(z / l)} \right]\;,\; p = \frac{1}{2}\rho \Omega 
^2\left[ {r^2 + \hat {P}(z / l)} \right]
\end{displaymath}

\noindent
where $l$ is some (as yet) unspecified length scale. The boundary conditions on 
$F$, $G$, $H$ and $\hat {P}$ are:
\begin{displaymath}
{\begin{array}{*{20}c}
 {z = 0\;\;:\,\;\;\;F = 0\;\;\;,\;\;\;G = 0\;\;,\;\;\;H = 0} \hfill \\
 {z \to \infty \;\;:\,\;\;\;F = 0\;\;\;,\;\;\;G = 1\;\;,\;\;\;\hat {P} = 0.} 
\hfill \\
\end{array} }
\end{displaymath}
\noindent
Substitution of the expression for \textbf{u} into (\ref{eq28}) yields three 
ordinary differential equations for $F$, $G$, $H$ and $\hat {P}$. However, from a 
physical point of view it is more interesting to work with (\ref{eq26}). First we 
note that the axial component of (\ref{eq26}) gives a differential equation for 
$\hat {P}$, from which we may deduce that $\hat {P}\sim \delta ^2$. In other 
words, $\hat {P}$ is a small perturbation in pressure within the boundary 
layer. Next we turn to the radial and azimuthal components of (\ref{eq26}). This, 
in turn, requires that we evaluate \textbf{J}~$\times $~\textbf{B}. From 
Ohm's law it is readily confirmed that
\begin{equation}
\label{eq29}
{\rm {\bf J}}\times {\rm {\bf B}} = \left[ { - \sigma u_r B^2,\;\; - \sigma 
B(u_\theta B - \partial V / \partial r),\,0} \right]
\end{equation}
\noindent
In order to fix $V$ we specify that there is no radial current, and hence no 
azimuthal Lorentz force, outside the boundary layer. In addition, (\ref{eq27}) 
demands
\begin{displaymath}
\nabla ^2V = \nabla \cdot \left( {{\rm {\bf u}}\times {\rm {\bf B}}} \right) 
= {\rm {\bf B}} \cdot \mathbf \omega 
\end{displaymath}

\noindent
from which we deduce that the electrostatic potential is of the form,
\begin{equation}
V = \frac{1}{2}B\Omega r^2 - 2B\Omega \int{\int{(1 - G)dzdz}} 
\end{equation}
\noindent
It follows that the Lorentz force is simply,
\begin{equation}
\label{eq30}
{\rm {\bf J}}\times {\rm {\bf B}} = \left[ { - \sigma B^2u_r ,\; - \sigma 
B^2(u_\theta - \Omega r),\;0} \right]
\end{equation}
\noindent
The radial and azimuthal components of (\ref{eq26}), along with 
the continuity equation,  then yield
\begin{eqnarray}
\label{eq31}
F^2 + H{F}' - G^2 + 1 = (\nu / \Omega l^2){F}'' - 2AF\\
\label{eq32}
2FG + {G}'H = (\nu / \Omega l^2){G}'' - 2A[G - 1]\\
{H}' + 2F = 0. \nonumber
\end{eqnarray}

\noindent
Finally, it is of interest to determine the 
magnitude of the current leaving the boundary layer. This is fixed by (\ref{eq27}) 
in the form,
\begin{equation}
\nabla \times {\rm {\bf J}} = \sigma ({\rm {\bf B}} \cdot \nabla ){\rm {\bf 
u}}
\end{equation}

\noindent
the axial component of which yields
\begin{equation}
\label{eq33}
(J_z )_\infty = 2\sigma B\Omega \int_0^\infty {(1 - G)dz} 
\end{equation}
\noindent
It is convenient to choose $l = \delta _\Omega $, the B\"{o}dewadt boundary 
layer thickness. Our governing equations for \textbf{u} then simplify to
\begin{eqnarray}
\label{eq34}
{F}'' = F^2 + H{F}' + (1 + G)(1 - G) + 2AF\\
\label{eq35}
{G}'' = 2FG + H{G}' + 2A(G - 1)\\
\label{eq36}
{H}' = - 2F,
\end{eqnarray}

These equations are readily solved numerically to give F, G and H. This then 
yields $(J_z )_\infty / \sigma B\Omega \delta _\Omega $ and $(u_z )_\infty / \Omega 
\delta _\Omega $ as functions of A, which is the primary information we need 
for the problems of sections 4 and 5. However, we shall see that it is 
possible to obtain analytical estimates of $J_{z}$ and $u_{z}$, which turn out 
to be more useful. 
\begin{figure}[htbp]
\centering
\includegraphics[width=0.75\textwidth]{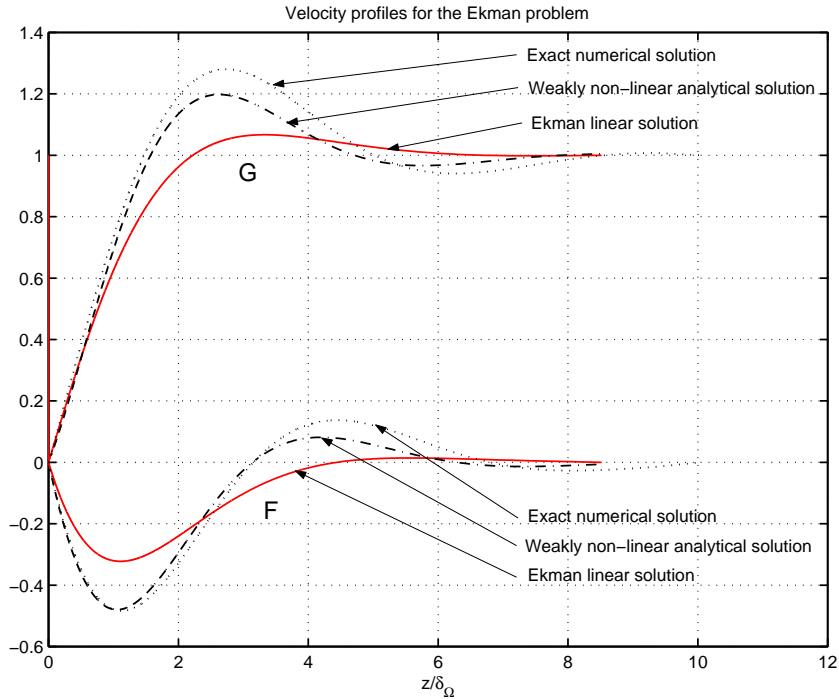}
\caption{B\"{o}dewadt layer profiles obtained under different assumptions. Solid: 
analytical solution of the Ekman problem. Dotted: numerical solution for the B\"odewadt problem 
calculated in section 4. Dashed: weakly non-linear solution. Curves above the $z-$ axis 
represent azimuthal velocity profiles and curves below represent radial velocity profiles.}
\label{ekman_sol}
\end{figure}
\noindent
When $A=0$ (no magnetic field) our governing equations represent the standard 
equations describing a B\"{o}dewadt layer. Let us consider this special case 
for a moment. The governing equations are,
\begin{eqnarray}
\label{eq37}
{F}'' = (1 + G)(1 - G) + F^2 + H{F}'\\
\label{eq38}
{G}'' = 2FG + {G}'H\\
\label{eq39}
{H}' = - 2F
\end{eqnarray}
\noindent

On integration these yield $F_\infty = 0$, $G_\infty = 1$ and $H_\infty = 
1.349$. Actually, as Greenspan points out, a rough approximation to the 
B\"{o}dewadt solution may be obtained by replacing the non-linear inertial 
terms on the right of (\ref{eq37}) and (\ref{eq38}) by the Coriolis force, $ - 2{\rm 
{\bf u}}_r \times \Omega $, where \textbf{u}$_{r}$ is the velocity measured 
in a frame of reference rotating with the fluid at infinity. This gives
\begin{equation}
\label{eq40}
{F}'' = 2(1 - G)\;\;\;,\;\;\;\;{G}'' = 2F
\end{equation}
\noindent
Of course, this leads to the Ekman solutions (\ref{eq19}), with $\Delta 
\Omega $ set equal to $\Omega $. Such a procedure is usually called the 
linear, or Ekman, approximation. Surprisingly, there is a reasonable 
qualitative agreement between the linear (Ekman) approximation and the exact 
non-linear solution (see \cite{green69}, and Figure \ref{ekman_sol}). In both cases the 
solution takes the form of a decaying oscillation of $F$ and $G-1$, and the 
frequency of oscillation is very similar in the two cases. However, the 
linear approximation over-estimates the exponential-like decay of $F$ and 
$G-1$ by a factor of about 2. It also underestimates $H_\infty $ by around 
35{\%}.\\

Turning now to the other extreme, of large $A$, equation (\ref{eq35}) reduces to 
\begin{equation}
\label{eq41}
{G}'' = 2A(G - 1)
\end{equation}
\noindent
This, plus (\ref{eq33}), yields
\begin{eqnarray}
\label{eq42}
(J_z )_\infty = 2\sigma B\Omega \delta _B \\
\label{eq43}
G = 1 - \exp \left[ { - z / \delta _B } \right]
\end{eqnarray}
\noindent
Of course, these coincide exactly with the results of the linear 
Ekman-Hartmann analysis (in the limit of large $A)$. The exact correspondence 
between (\ref{eq43}) and (\ref{eq22}), when $\Delta \Omega $ is set equal to $\Omega $, 
is inevitable since in both cases the non-linear inertial terms are 
neglected.

The general picture, then, is that the linear Ekman-Hartmann approximation 
(with $\Delta \Omega $ set equal to $\Omega )$ yields results which are 
qualitatively similar to the B\"{o}dewadt-Hartmann problem when $A$~=~0, and 
that the two analyses coincide when $A$ becomes large. We now show how to 
obtain an improved approximation to the B\"{o}dewadt-Hartmann solution which 
has a simple algebraic form. We follow a method originally developped for B\"odewadt 
layers (see for example \cite{cole68}).

\subsection{An Approximate Analytical Solution}
Before tackling the weakly non-linear problem, it is important to note that 
the full system (\ref{eq34})-(\ref{eq35}) with associated boundary conditions need not have 
a unique solution (see \cite{zandbergen87}). Physically, however this most likely 
relates to the absence of lateral boundary 
conditions, which appear to play a determining role in real experiments. The solution we 
look at is the physically most important, in that it is the one which appears in practice when fixed 
lateral boundaries are included at large radius.   

Let us return to (\ref{eq34}) and (\ref{eq35}) and look for solutions at large $z$. If we 
linearise the equations around the far-field solution $(F,G,H) = 
(0,1,H_\infty )$ we obtain
\begin{eqnarray}
\label{eq44}
{F}'' - H_\infty {F}' - 2AF = - 2\hat {G}\\
\label{eq45}
{\hat {G}}'' - H_\infty {\hat {G}}' - 2A\hat {G} = 2F
\end{eqnarray}

\noindent
where $\hat {G} = G - 1$. This yields oscillatory solutions of the form,
\begin{equation}
\label{eq46}
\hat {G}_\infty ,F_\infty \sim \exp \left[ { - (\hat {R} - H_\infty / 2)z / 
\delta _\Omega } \right]\exp \left[ {\pm jz / \hat {R}\delta _\Omega } 
\right]
\end{equation}

\noindent
where
\begin{equation}
\label{eq47}
\hat{R} = \left[ {\hat{A}+(1+\hat{A}^2)^{1/2}} \right]^{1/2}
, \quad \hat {A} = A + H_\infty ^2 / 8, \quad j^2 = -1
\end{equation}
\noindent
Note that if we set $H_\infty $ to zero in (\ref{eq46}) and (\ref{eq47}) we obtain the 
linear Ekman estimate. Let us now make two approximations. First, we take 
$H_\infty $ to be given by the linear Ekman-Hartmann solution (\ref{eq17}):
\begin{equation}
H_\infty = \frac{2R}{1 + R^4}\;\;\;\;,\;\;\;\;\;R = \left[ {A + \left( {1 + 
A^2} \right)^{1 \mathord{\left/ {\vphantom {1 2}} \right. 
\kern-\nulldelimiterspace} 2}} \right]^{1 \mathord{\left/ {\vphantom {1 2}} 
\right. \kern-\nulldelimiterspace} 2}
\end{equation}
\noindent
Second, we assume that our estimate of $\hat {G}_\infty $ and $F_\infty $ 
are valid, not just for large $z$, but for all $z$. If this is true then,
\begin{eqnarray}
\label{eq48}
\hat {G} = G - 1 = - \exp \left[ { - (\hat {R} - H_\infty / 2)z / \delta 
_\Omega } \right]\cos (z / \hat {R}\delta _\Omega )\\
\label{eq49}
F = - \exp \left[ { - (\hat {R} - H_\infty / 2)z / \delta _\Omega } 
\right]\sin (z / \hat {R}\delta _\Omega )
\end{eqnarray}
\noindent
Let us now see how our guesses have faired. We look first at small $A$. When 
$A$~=~0 (a pure B\"{o}dewadt layer) we have $\hat {R} = 1.064$ and the 
resulting curves for $F$ and $G$ are plotted in Figure 5. The exact solution and 
the linear Ekman approximation are also given for comparison. Evidently, 
there is a reasonable correspondence between (\ref{eq48}) and (\ref{eq49}) and the exact 
solution. For large $A$, on the other hand, (\ref{eq48}) and (\ref{eq49}) reduce to
\begin{equation}
\label{eq50}
F = 0\,\,\,,\,\,\,G = 1 - \exp \left[ { - z / \delta _B } \right]
\end{equation}

\noindent
which corresponds precisely with both the exact solution and the Ekman 
approximation.
 
Given that (\ref{eq48}) and (\ref{eq49}) are reasonably accurate for small $A$, and exact 
for large $A$, our proposal is to adopt them as approximations to the 
B\"{o}dewadt-Hartmann layer in sections 4 and 5. The corresponding current 
distribution is given by (\ref{eq33}) and this, combined with (\ref{eq48}), fixes 
$J_{z}$. In summary then, we have the following estimates of $(u_z )_\infty $ 
and $(J_z )_\infty $:
\begin{eqnarray}
\label{eq51}
\frac{(u_z )_\infty }{\Omega \delta _\Omega } = \frac{2R}{1 + 
R^4}\;\;\;\;,\;\;\;\;R = \left[ {A + (1 + A^2)^{1 \mathord{\left/ {\vphantom 
{1 2}} \right. \kern-\nulldelimiterspace} 2}} \right]^{{\,\;1} 
\mathord{\left/ {\vphantom {{\,\;1} 2}} \right. \kern-\nulldelimiterspace} 
2}\\
\label{eq52}
\frac{(J_z )_\infty }{2\sigma B\Omega \delta _\Omega } = \frac{\hat 
{R}^2(\hat {R} - H_\infty / 2)}{1 + \hat {R}^2(\hat {R} - H_\infty / 
2)^2}\;\;\;\;,\;\;\;\;\hat {R} = \left[ {\hat {A} + (1 + \hat {A}^2)^{1 
\mathord{\left/ {\vphantom {1 2}} \right. \kern-\nulldelimiterspace} 2}} 
\right]^{\;1 \mathord{\left/ {\vphantom {1 2}} \right. 
\kern-\nulldelimiterspace} 2}
\end{eqnarray}

\subsection{Numerical Solutions of the Governing Equations}

Before adopting (\ref{eq51}) and (\ref{eq52}) it seems sensible to compare these with 
the exact solutions of (\ref{eq34})-(\ref{eq36}) obtained by numerical means. 
First of all, the problem is expressed on a finite interval, using the 
variable shift $\frac{z}{\delta _\Omega } = - \ln \left( {1 - t} \right)$. 
The resulting system is then discretized with a centred finite differences 
method. The associated non-linear set of equations is solved using a 
Newton-Raphson algorithm (the equivalent first-order system is then 
5-dimensional). In order to be able to compute the large number of points 
required to reach high values of $z$ in a reasonable computation time, we need a 
fast matrix inversion. We proceed as follows: the $5n$ equations (where $n$ is the 
number of points in the mesh) are ordered so that the finite 
difference system is represented by a bi-diagonal 5$\times $5-block matrix 
(diagonal blocks $J_{k}$ and sub-diagonal blocks $K_{k}$ ) with a block C in 
the upper right corner containing the boundary conditions:
\begin{equation}
\left[ {{\begin{array}{*{20}c}
 {J_1 } \hfill & \hfill & \hfill & C \hfill \\
 {K_2 } \hfill & {J_2 } \hfill & \hfill & \hfill \\
 \hfill & {...} \hfill & {...} \hfill & \hfill \\
 \hfill & \hfill & {K_n } \hfill & {J_n } \hfill \\
\end{array} }} \right]\times \left[ {{\begin{array}{*{20}c}
 {X_1 } \hfill \\
 {X_2 } \hfill \\
 {...} \hfill \\
 {X_n } \hfill \\
\end{array} }} \right] = \left[ {{\begin{array}{*{20}c}
 {F_1 } \hfill \\
 {F_2 } \hfill \\
 {...} \hfill \\
 {F_n } \hfill \\
\end{array} }} \right]
\end{equation}

A first system S$_{1}$ is formed with the first block-line. It involves 
unknown blocks $X_{1 }$and $X_{n }$ (the unknown vector $X$ is split into $n$ 
5-dimensional ``block-vectors''). The unknown $X_{n}$ is expressed 
recursively as a function of $J_{k}$, $K_{k}$ \textit{/ k$ \in ${\{}2..n{\}}} and $X_{1}$ thanks to the 
bi-diagonal structure of the matrix. The resulting system of 5 equations 
($i.e.$ one block) can then be added to S$_{1}$ to give an invertible system whose 
solutions are $X_{1}$ and $X_{n}$. The other unknowns are then deduced 
recursively. This inversion method requires a number 
of operations proportional to $n$ (versus $n^2$, if the Jacobian matrix had 
been directly inverted) which considerably reduces the computation time. 
The accuracy of the procedure was checked by comparing the analytical and numerical
 solutions of the Ekman problem.\\
\indent
We first investigate the non-magnetic case ($A=0$). Figure \ref{ekman_sol} shows a comparison 
between different estimates the B\"odewadt problem: the fully non-linear solution of 
(\ref{eq37})-(\ref{eq39}) (obtained numerically 
on 23000 points), the weakly non-linear solution (\ref{eq48})-(\ref{eq49}) and the 
 linear Ekman approximation. It appears 
that non-linear effects are responsible for rather stronger oscillations in the 
velocity profiles than those predicted by (\ref{eq48})-(\ref{eq49}). This is consistent with the assumptions on which the 
analytical solution (\ref{eq48})-(\ref{eq49}) relies, as the latter extrapolates a 
solution (\ref{eq46})-(\ref{eq47}) valid for large $z$ and takes the value returned by the 
linear Ekman-Hartmann theory for $H_{\infty }$. The associated Ekman pumping is 
then underestimated by \textit{35{\%}} (compared to the fully non-linear solution), and so 
are the radial velocity and the oscillations. For $A=0$, the discrepancy between 
(\ref{eq48})-(\ref{eq49}) and the full numerical result is below \textit{10{\%}} on azimuthal and 
radial velocity, which is not such an expensive price to pay for an 
analytical solution.

\begin{figure}
\centering
\includegraphics[width=0.75\textwidth]{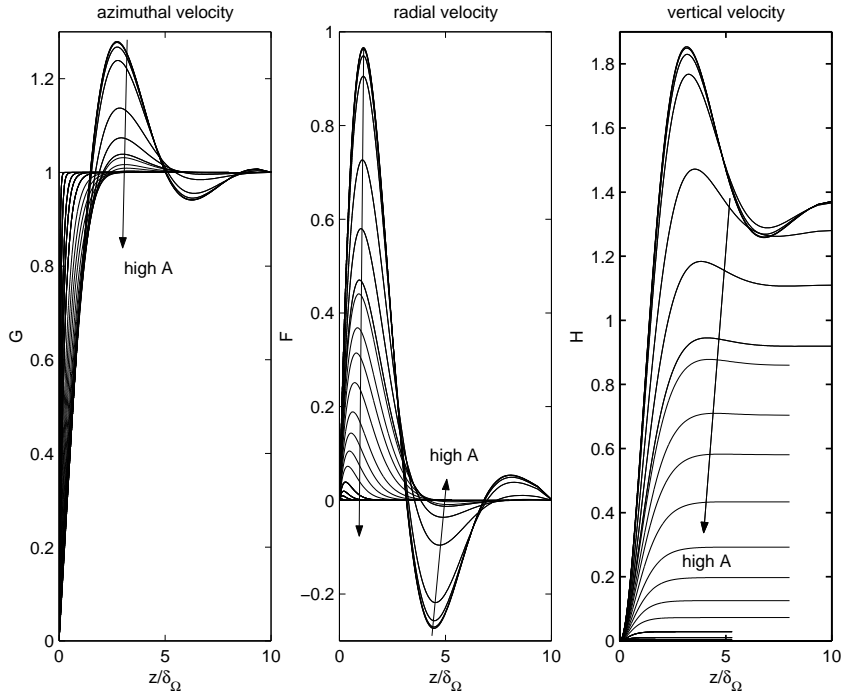}
\caption{Velocity profiles in the B\"odewadt-Hartmann layer for values of $A$
in the range $10^{-3} \to 10^3$. The arrows go from low to high $A$ curves.}
\label{vitesses}
\end{figure}

\begin{figure}
\centering
\includegraphics[width=0.75\textwidth]{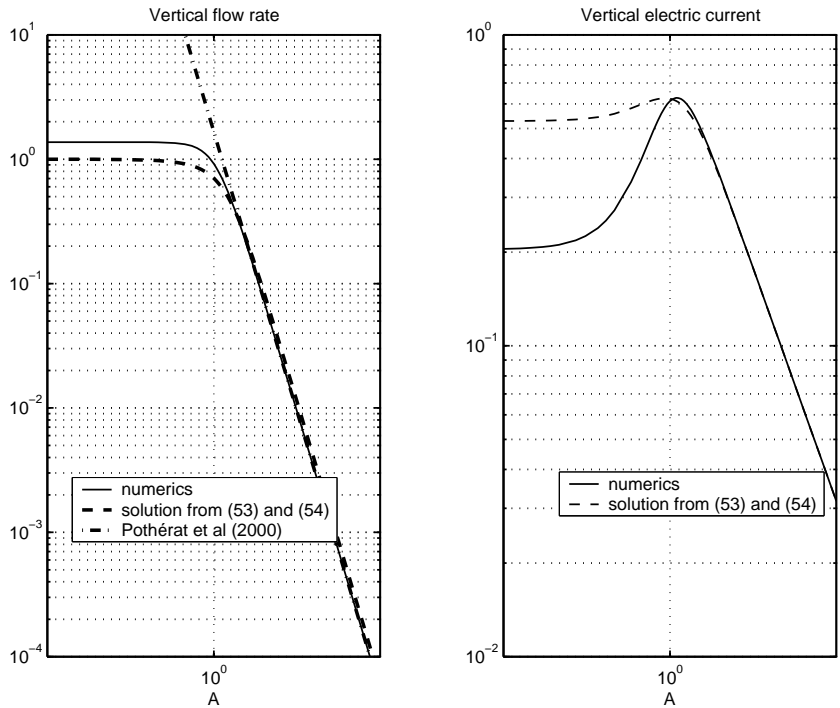}
\caption{Vertical velocity (left) and vertical electric current (right) at the edge 
of the B\"odewadt-Hartmann layer. The numerical simulation is given by the solid line 
and the approximate solution (\ref{eq51}-\ref{eq52}) by the dashed line.}
\label{debits}
\end{figure}

We now turn our attention toward the non-linear MHD case (\textit{A$ \ne $}0). System 
(\ref{eq37}-\ref{eq39}) is solved numerically for values of $A$ taken in the range $10^{ - 
3} \rightarrow 10^{3}$ (figure \ref{debits}). 
For A$<1$, the velocity 
profiles across the layer are close to the well-known B\"odewadt layer 
profile, but with the difference that the oscillating part of the profile is 
softened by the action of the magnetic field, as predicted by (\ref{eq48})-(\ref{eq49}). 
For $A>>1,$ the profile is rather close to the exponential profile of the Hartmann 
layer without oscillations. In this case, the results of the numerical 
simulation may be compared not only with our approximate solution (\ref{eq48})-(\ref{eq49}),
 but also with those of \cite{psm00}, which applies to Hartmann layers with weak inertia.
 Figure \ref{debits} 
 shows the various estimates of $(u_z)_\infty$ and $(J_z)_\infty$, the vertical velocity 
 and current leaving the boundary layer. The left-hand figure shows $(u_z)_\infty$. It can be seen 
 that our approximate solution (\ref{eq51}) is close to the exact value for all $A$. The  model 
 of \cite{psm00} is accurate for $A>2$ but not for small $A$. The right-hand figure shows $(J_z)_\infty$ 
 corresponding to approximation (\ref{eq52}), as well as the exact, computed value. The approximate solution 
 is good for $A>1$ but overestimates the current for $A<1$.\\
Note that \cite{desjardins99}, \cite{desjardins01} have investigated the stability of 
geophysical Ekman-Hartmann layers. Their study differs from the present 
problem by the geometry (spherical) and also by the fact that magnetic field 
and rotation are not aligned nor orthogonal to the layer. The results 
however do not depend on the co-latitude (along which the angle between the 
layer and the rotation vary) which suggest that they might be of some 
relevance here. In the non-magnetic case, it is found that the flow is 
non-linearly unstable for values of the Reynolds number scaled on the boundary 
layer thickness above 0.55 ($A=0)$ and linearly unstable for values above 40. The 
presence of the magnetic field makes the flow more stable so that these 
values are changed to 0.71 and 160 respectively for $A=1$. (The linear stability 
value corresponds to the co-latitude for which the rotation is orthogonal to 
the boundary layer). This is consistent with the fact that plane Hartmann 
layers are indeed much more stable than rotation layers. (They have a
 linear stability threshold around 50000, according to 
\cite{roberts67}). These results underline the fact that 
the solutions obtained numerically in this section are only valid below a 
threshold value of $\Omega$, which increases with $A$. For stronger rotation, 
\cite{desjardins01} showed that traveling waves appear in the plane of the 
layer.

\section{Forced Vortex in a Confined Layer}

We now look at flows which are typical of laboratory experiments. In particular, 
we consider a pool of depth $W$, the depth being 
assumed to be much greater than the B\"{o}dewadt-Hartmann boundary layer 
thickness (figure \ref{fig3}). That is, we restrict ourselves to free surface 
flows which have a 
high Reynolds number. There are two particular cases of interest. The first 
is where a steady vortex is maintained by an external azimuthal force, say 
that produced by a rotating magnetic field. We shall study that problem 
here. The second, which we leave to section 5, is the transient problem of 
spin down from some initial state of rotation. The geometry for both cases 
is shown in Figure \ref{fig3}. For simplicity, we model the free surface at
$z=W$ as a symmetry plane.

In this section we look at the case where the vortex is maintained by the 
body force,
\begin{equation}
{\rm {\bf F}} = \frac{1}{2}\Omega _f r{\rm {\bf \hat {e}}}_\theta 
\;\;\;\;\;\;\;\;\;\;,\;\;\Omega _f = constant
\end{equation} 
\noindent
Since $F_\theta $ is linear in $r$ we can once again look for solutions of the 
Karman type. The resulting equations are of a form similar to (\ref{eq29})-(\ref{eq33}). 
That is, if we look for solutions of the form 
\begin{equation}
\label{eq53}
{\rm {\bf u}} = \left[ {\Omega _c rF\left( {z / l} \right)\;\;,\;\;\;\Omega 
_c rG(z / l)\;\;,\;\;\;\Omega _c lH(z / l)} \right]
\end{equation}

\noindent
we find,
\begin{eqnarray}
\label{eq54}
{H}' + 2F = 0\\
\label{eq55}
F^2 + H{F}' - G^2 + 1 = (\nu / \Omega _c l^2){F}'' - 2AF\\
\label{eq56}
2FG + {G}'H = (\nu / \Omega _c l^2){G}'' + \frac{1}{2}\left( {\Omega _f / 
\Omega _c } \right)^2 - 2A\left[ {G - f} \right]
\end{eqnarray}
\noindent
Here $\Omega _c $ is a typical rotation rate outside the boundary layer, $l$ is 
an arbitrary length scale, and $f$ is related to the radial gradient of $V$,
\begin{displaymath}
f = \left( {\Omega _c rB} \right)^{ - 1}\frac{\partial V}{\partial r}
\end{displaymath}
\noindent
(See equation (\ref{eq29}) for the origin of the term $G-f$.) The only differences 
between (\ref{eq54})-(\ref{eq56}) and (\ref{eq31})-(\ref{eq32}) is that: (i) we have incorporated 
the driving force $\frac{1}{2}\Omega _f^2 r$ ; and (ii) we have yet to 
specify $f$. In the semi-infinite domain problem of section 3 we specified that 
$J_r $ is zero outside the boundary layer and this fixed $f$ as $f = 1$. 
However, it is clear from Figure \ref{fig3} that this is no longer legitimate.
\noindent
Now we know that there is a uniform flux of current out of the boundary 
layer, which we called $(J_z )_\infty $. Thus the poloidal current in the 
core satisfies
\begin{eqnarray}
\label{eq57}
\nabla \cdot {\rm {\bf J}}_p = 0\,\,\,\,\,,\,\,\,\,\,\,\nabla \times {\rm 
{\bf J}}_p = \sigma {\rm {\bf B}} \cdot \nabla {\rm {\bf u}}_\theta \\
\label{eq58}
J_z = 0\;\;\;\mbox{on }z = W\;\;,\;\;\;J_z \to (J_z )_\infty \mbox{ as }z 
\to 0.
\end{eqnarray}
\noindent
We shall see shortly that $u_\theta $ is virtually independent of $z$ in the 
core and so (\ref{eq57}) and (\ref{eq58}) have the unique solution:
\begin{equation}
\label{eq59}
(J_r )_c = \frac{(J_z )_\infty }{2}\,\frac{r}{W}\,\;\;\;,\;\;\;\;(J_z )_c = 
(J_z )_\infty \left[ {1 - \frac{z}{W}} \right]
\end{equation}
\noindent
This represents an outward flow of current in the core, as indicated in 
Figure 3. From (\ref{eq59}) we can find $V$ and it follows that, in the core of the 
flow,
\begin{equation}
\label{eq60}
f_c = G_c - \frac{(J_z )_\infty }{2\sigma \Omega _c WB}
\end{equation}
\noindent
We may simplify (\ref{eq60}) using (\ref{eq33}) in the form
\begin{equation}
\label{eq61}
(J_z )_\infty = 2\sigma B\Omega _c \int_0^\delta {(1 - G)dz} 
\end{equation}

\noindent
which gives,
\begin{equation}
\label{eq62}
f_c = G_c - W^{ - 1}\int_0^\delta {(1 - G)dz} 
\end{equation}
\noindent
Thus 1~-~$f_{c}$ is of the order of $\delta / W$ in the core of the flow. In 
the boundary layer, on the other hand, we may continue to take $f$~=~1 since 
the curvature of the current lines in the core is negligible on the scale of 
\textit{$\delta $}. 
\noindent
We are now in a position to write down the governing equations for the core 
and for the boundary layer. In the boundary layer we take $l = \delta 
_\Omega $, which gives
\begin{eqnarray}
\label{eq63}
{F}''_b = F_b^2 + H_b {F}'_b - G_b^2 + 1 + 2AF_b \\
\label{eq64}
{G}''_b = 2F_b G_b + H_b {G}'_b + 2A(G_b - 1) - \frac{1}{2}\left( {\Omega _f 
/ \Omega _c } \right)^2\\
\label{eq65}
{H}'_b = - 2F_b 
\end{eqnarray}
\noindent
These are identical to the equations of Section 3 except for the forcing 
term. (Consult equations (\ref{eq34})-(\ref{eq36}).) In the core, on the other hand, we 
take $l = W$ and neglect the viscous stresses since $\Omega W^2 / \nu > > 1$. 
The result is
\begin{eqnarray}
\label{eq66}
{H}'_c + 2F_c = 0\\
\label{eq67}
F_c^2 + H_c {F}'_c - G_c^2 + 1 = - 2AF_c \\
\label{eq68}
2F_c G_c + {G}'_c H_c = \frac{1}{2}\left( {\Omega _f / \Omega _c } \right)^2 
- 2A\left( {G_c - f_c } \right)
\end{eqnarray}
\noindent
We now introduce the parameter $\varepsilon = \delta _\Omega / W$. Recall 
that we consider $Re = \Omega W^2 / \nu$ to be asymptotically large but retain 
$A$ as zero or finite. It follows that $\varepsilon \to 0$ as $\nu \to 0$ 
irrespective of the value of $A$. Now the matching condition on $u_z $ at the 
edge of the boundary layer gives
\begin{displaymath}
H_c (z_c \to 0) = \varepsilon H_b (z_b \to \infty ) = \varepsilon (H_b 
)_\infty 
\end{displaymath}

\noindent
where $z_c = z / W$ and $z_b = z / \delta _\Omega $. It follows that $H_c $ 
and $F_c $ are both of order $\varepsilon $. We now expand $H_{c}$, $F_{c}$ 
and $G_{c}$ in powers of $\varepsilon $ and look for solution of (\ref{eq66}) - 
(\ref{eq68}). We find that \textbf{u} has the same structure as \textbf{J} in the 
core:
\begin{eqnarray}
\label{eq69}
F_c = \frac{\varepsilon }{2}\left( {H_b } \right)_\infty + 0(\varepsilon )\\
\label{eq70}
G_c = 1 + 0(\varepsilon)\\
\label{eq71}
H_c = \varepsilon (H_b )_\infty \left[ {1 - z / W} \right] + 0(\varepsilon)
\end{eqnarray}
\noindent
It follows that the azimuthal equation of motion reduces to
\begin{equation}
\label{eq72}
\varepsilon (H_b )_\infty + 2AW^{ - 1}\int_0^\delta {(1 - G_b )dz = 
\frac{1}{2}(\Omega _f / \Omega _c )^2} 
\end{equation}
\noindent
If we retrace our steps to find the origin of these terms we discover that 
(\ref{eq72}) is simply a statement of
\begin{equation}
\label{eq73}
2u_r \Omega _c + \rho ^{ - 1}J_r B = F_\theta 
\end{equation}
\noindent
It appears that $F_\theta $, is balanced either by the Coriolis force, 
$2{\rm {\bf u}}\times \Omega $, or else the Lorentz force, $\rho ^{ - 1}{\rm 
{\bf J}}\times {\rm {\bf B}}$. Thus, as noted in Section 1, the dynamics of 
the core is determined by the radial components of ${\rm {\bf u}}_c $ and 
${\rm {\bf J}}_c $. These, in turn, depend on the axial flux of current and 
mass released by the B\"{o}dewadt-Hartmann layer.

Let us now turn to the boundary-layer equations. From (\ref{eq72}) we see that 
$\frac{1}{2}\left( {\Omega _f / \Omega _c } \right)^2$ is of order 
$\varepsilon H_b $ and so the forcing term in (\ref{eq64}) is negligible. The 
dynamical equations for the boundary therefore reduce to those of Section 3. 
It follows, from (\ref{eq51}) and (\ref{eq52}), that,
\begin{displaymath}
(H_b )_\infty = \frac{2R}{1 + R^4},\;\;\;\;\;\;\;\;\;\;\;\;\;\;\;\;R = 
\left[ {A + (1 + A^2)^{1/2}}\right]^{1/2}
\end{displaymath}
\begin{displaymath}
\frac{1}{\delta _\Omega }\int_0^\delta {(1 - G_b )dz = \frac{\hat {R}^2(\hat 
{R} - H_\infty / 2)}{1 + \hat {R}^2(\hat {R} - H_\infty / 2)^2}} 
\,\,\,,\,\,\,\hat {R} = \left[ {\hat {A} + (1 + \hat {A}^2)^{1/2}} 
\right]\,^{1 / 2}
\end{displaymath}
\noindent
The azimuthal force balance in the core therefore reduces to
\begin{equation}
\label{eq74}
 \underbrace{\frac{2R}{1 + R^4}}_{Coriolis} + \underbrace{2A\frac{\hat {R}^2(\hat {R} - H_\infty / 2)}{1 + \hat 
{R}^2(\hat {R} - H_\infty / 2)^2}}_{Lorentz} = \frac{W}{2\delta _\Omega }\left( 
{\frac{\Omega _f }{\Omega _c }} \right)^2  
\end{equation}
\noindent
When $A$ is small (negligible magnetic field) we have a balance between the 
Coriolis force and $F_\theta $, which yields, 
\begin{equation}
\label{eq75}
\Omega _c = \frac{\Omega _f }{2^{2/3}}\;\left[ {\frac{\Omega _f W^2}{\nu}} \right]^{1 
/3} (A\to 0)
\end{equation}
\noindent
This result was first obtained by \cite{dav92}. In the event that the 
Coriolis force is negligible, on the other hand, we find,
\begin{equation}
\label{eq77}
\Omega _c = \frac{\Omega _f^2 W\delta _B }{2\nu}
\end{equation}
\noindent
However, it is unlikely that we can ever reach a situation in which the 
Coriolis force is negligible. To see why this is so, we must rewrite (\ref{eq74}) 
in a way in which (the undetermined) $\Omega _c $ is made more explicit.
Let $A_f = \left( {2\tau \Omega _f } \right)^{ - 1}$, $(Re)_f = \Omega _f 
W^2 / \nu$ and $\lambda = \Omega _f \left( {Re} \right)_f^{1 \mathord{\left/ 
{\vphantom {1 3}} \right. \kern-\nulldelimiterspace} 3} / \Omega _c $. Then 
our force balance (\ref{eq74}) becomes
\begin{displaymath}
\frac{2R}{1 + R^4} + 2A_f \frac{\hat {R}^2\left( {\hat {R} - H_\infty / 2} 
\right)}{1 + \hat {R}^2\left( {\hat {R} - H_\infty / 2} \right)^2}\left( 
{Re} \right)_f^{ - 1 \mathord{\left/ {\vphantom {1 3}} \right. 
\kern-\nulldelimiterspace} 3} \lambda = \frac{1}{2}\lambda ^{3 
\mathord{\left/ {\vphantom {3 2}} \right. \kern-\nulldelimiterspace} 2}
\end{displaymath}
\noindent
We now let $(Re)_f \to \infty $ while retaining $A_{f}$ as finite. The 
Lorentz term then goes to zero and we are left with a balance between the 
Coriolis force and $F_{\theta }$. The estimate of \textit{$\Omega $}$_{c}$ is then
\begin{equation}
\label{eq78}
\Omega _c = \left({\frac{4R}{1 + R^4}}\right)^{-2/3}(Re)_f^{1/3} 
\Omega _f \;\;\;\;\; (\textit{Any A})
\end{equation}

\section{Spin-down from Some Initial State}
\indent
It is well-known that Karman-type similarity also extends to unsteady flows 
(see, for example, \cite{green69}). It is necessary only to replace the 
forcing term, $\frac{1}{2}\Omega _f^2 r$, in the azimuthal equation (\ref{eq56}) by 
${ - \partial u_\theta } \mathord{\left/ {\vphantom {{ - \partial u_\theta } 
{\partial t}}} \right. \kern-\nulldelimiterspace} {\partial t}$. There is 
also a deceleration term $ - {\partial u_r } \mathord{\left/ {\vphantom 
{{\partial u_r } {\partial t}}} \right. \kern-\nulldelimiterspace} {\partial 
t}$ on the right of (\ref{eq55}). However, this turns out to be negligible by 
comparison with the other inertial terms, essentially because the spin-down 
time is relatively slow. We now repeat all of the steps leading up to (\ref{eq72}) 
and find,
\begin{equation}
\label{eq79}
\varepsilon (H_b )_\infty + 2AW^{ - 1}\int_0^\delta {(1 - G_b )dz = - \Omega 
_c^{ - 2} \partial \Omega _c / \partial t} 
\end{equation}
\noindent
Physically, this represents the balance,
\begin{displaymath}
2u_r \Omega _c + \rho ^{ - 1}J_r B = - \partial u_\theta / \partial t
\end{displaymath}
\noindent
Since we are now considering an initial value problem it is convenient to 
introduce $\hat {t} = \Omega _0 t$, $\hat {\Omega } = \Omega / \Omega _0 $, 
$\varepsilon _0 = (Re)_0^{ - 1 \mathord{\left/ {\vphantom {1 2}} \right. 
\kern-\nulldelimiterspace} 2} = (\nu / \Omega _0 W^2)^{1 \mathord{\left/ 
{\vphantom {1 2}} \right. \kern-\nulldelimiterspace} 2}$ and $A_0 = (2\Omega 
_0 \tau )^{-1}$, where $\Omega _0 = \Omega (t = 0)$. Our force balance can 
then be rewritten as, 
\begin{equation}
\label{eq80}
\varepsilon _0 \hat {\Omega }^{3 \mathord{\left/ {\vphantom {3 2}} \right. 
\kern-\nulldelimiterspace} 2}(H_b )_\infty + \hat {\Omega }2A_0 W^{ - 
1}\int_0^\delta {(1 - G_b )dz = - \partial \hat {\Omega } / \partial \hat 
{t}} 
\end{equation}
\noindent
Substituting now for $(H_b )_\infty $ and $\int{\left( {1 - G_b }
\right)dz} $ using (\ref{eq51}) and (\ref{eq52}) yields,
\begin{equation}
\label{eq81}
\varepsilon _0 \hat {\Omega }^{3/2} 
\frac{2R}{1 + R^4} + 2A_0 \varepsilon _0 \hat 
{\Omega }^{1/2}\frac{\hat {R}^2(\hat {R} - H_\infty / 2)}{1 + 
\hat {R}^2(\hat {R} - H_\infty / 2)^2} = - \frac{\partial \hat {\Omega 
}}{\partial t}
\end{equation}
\noindent
This is too complicated to integrate by analytical means for the general case because $R$ and $\hat 
{R}$ are themselves functions of $\Omega $. However, we can integrate (\ref{eq81}) 
for the two extremes of $A_0 \to 0$, $A_0 \to \infty $. When $A_0 = 0$ we 
have $R$~=~1 and (\ref{eq81}) simplifies to
\begin{equation}
\label{eq82}
\frac{\partial \hat {\Omega }}{\partial \hat {t}} + \varepsilon _0 \hat 
{\Omega }^{3 \mathord{\left/ {\vphantom {3 2}} \right. 
\kern-\nulldelimiterspace} 2} = 0
\end{equation}
\noindent
This integrates to give
\begin{equation}
\label{eq83}
\begin{array}{l}
 \Omega / \Omega _0 = \left( {1 + \frac{t}{t_E }} \right)^{ - 2} \\ 
 t_E = \left( {\frac{1}{2}\varepsilon _0 \Omega _0 } \right)^{ - 1} \\ 
 \end{array},
\end{equation}

\noindent
where $t_{E }$ is the typical friction time associated to a linear Ekman 
boundary layer. On the other hand, when $A_{0}$ is very 
large ($i.e.$ for negligible inertia), $R = 
\hat {R} = \left( {2A} \right)^{1 \mathord{\left/ {\vphantom {1 2}} \right. 
\kern-\nulldelimiterspace} 2}$ and (\ref{eq81}) reduces to
\begin{equation}
\label{eq84}
\frac{\partial \hat {\Omega }}{\partial \hat {t}} + \left( {2A_0 \varepsilon 
_0^2 } \right)^{1 \mathord{\left/ {\vphantom {1 2}} \right. 
\kern-\nulldelimiterspace} 2}\hat {\Omega } = 0
\end{equation}

\noindent
which yields
\begin{equation}
\label{eq85}
\begin{array}{l}
 \Omega = \Omega _0 \exp \left[ { - t / t_H } \right] \\ 
 t_H = \frac{W\delta _B }{\nu } \\ 
 \end{array}
\end{equation}

\noindent
where $t_{H}$ is a time which characterizes Hartmann layer friction.

\begin{figure}[htbp]
\begin{center}
\includegraphics[width=0.75\textwidth]{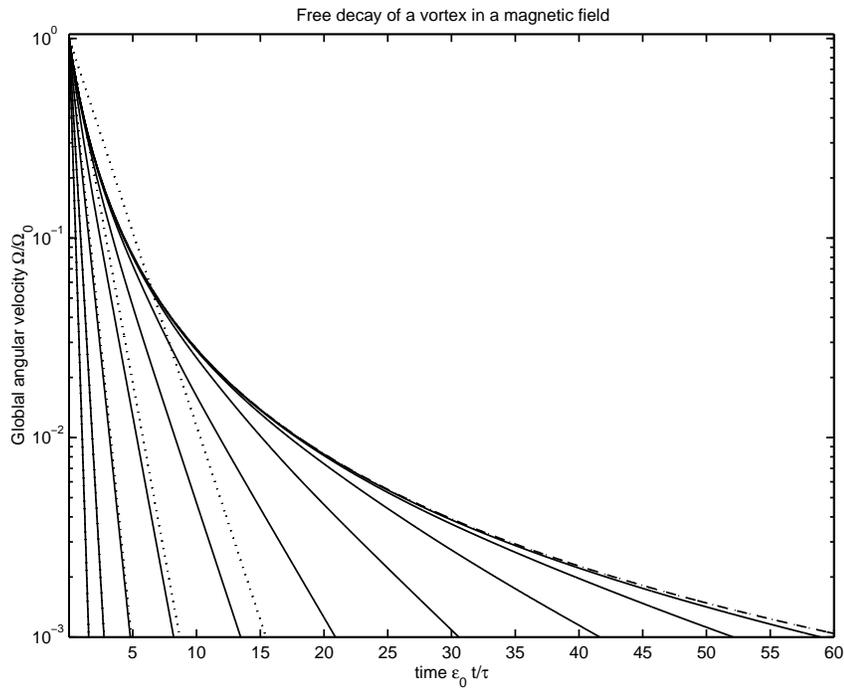}
\caption{Free decay of a vortex in a magnetic field. Dash-doted lines: Bodewadt 
case ($A_{0}=0)$. Solid: numerical solution of (\ref{eq81}) for Elsasser number equal to 
{\{}0.00032, 0.001, 0.0032, 0.01, 0.032, 0.1, 0.32, 1, 3.16, 10 {\}} 
(stronger decay when $A_{0}$ increases). Dotted lines: linear approximation 
(\ref{eq85}) for $A_{0}$ in {\{}0.1, 0.32, 3.16, 10{\}}. The last two dotted lines 
cannot be distinguished from the fully non-linear solution.}
\label{decay}
\end{center}
\end{figure}

The general equation (\ref{eq81}) has been solved numerically for values of the 
initial Elsasser number in the range $10^{-3} \to 10^3$ (see figure \ref{decay}). The approximation (\ref{eq85}) 
turns out to be very accurate when $A_{0}$\textit{$ \geqslant $1. } Even for an initially dominant 
rotation, the decay becomes exponential when $A$ reaches a value of the order of 
$1$. 

\section{Conclusions}

The first part of this work has provided some weakly non-linear analytical 
solutions to the semi-infinite Hartmann-B\"{o}dewadt problem, which provides 
a better approximation than the usual Ekman linear approximation. These new 
velocity profiles turn out to be quite close to the fully non-linear 
numerical solution (though they slightly underestimate the oscillating part 
of the profile), which justifies their use in further work. The numerical 
results are new as well, and together with the analytical solutions, they 
point out that the B\"{o}dewadt-Hartmann layer becomes very close to the 
simple Hartmann layer exponential profile as soon as the Elsasser number 
reaches a few units. 
The second part (sections \textbf{4} and \textbf{5}) of this work tackles 
the problem of a forced or free vortex in a confined layer of fluid, in which the 
B\"{o}dewadt-Hartmann boundary layer is shown to have the same 
dynamics as in the semi-infinite problem. As the core flow directly depends 
on the quantities injected by the boundary layer into the core (vertical flow 
rate and vertical electric current), the results of the semi-infinite 
problem allows us to derive an expression for the core global angular 
velocity both in the case of a constant forcing and for the spin-down from 
some initial value of the rotation. In the latter case, it is found that 
meridian electric current and secondary flows essentially result in effects 
similar to friction, with a characteristic time varying from a linear Ekman 
layer characteristic friction time (when rotation dominates electromagnetic 
effects) to the Hartmann layer friction time (when electromagnetic 
effects dominates rotation).

\textit{Acknowledgements}

The authors would like to thank Ren\'{e} Moreau and Jo\"{e}l 
Sommeria for the fruitful discussions on this work.




\end{document}